\documentclass[lettersize,journal]{IEEEtran}%change 1
\usepackage{amsmath,amsfonts}
\usepackage{algorithmic}
\usepackage{algorithm}
\usepackage{array}
\usepackage[caption=false,font=normalsize,labelfont=sf,textfont=sf]{subfig}
\usepackage{textcomp}
\usepackage{stfloats}
\usepackage{url}
\usepackage{verbatim}
\usepackage{graphicx}
\usepackage{cite} % Nga uncomment in March 14th 2025 for reference problem on ArXiv
\usepackage{cite}  % This ensures proper IEEE-style citations

\usepackage{silence}

\usepackage{lipsum} % for dummy text
\usepackage{url}
\usepackage{xcolor}
\usepackage{hyperref}
\usepackage{cleveref}
\usepackage{siunitx}

\usepackage{soul}

\usepackage{xspace}

\definecolor{darkgreen}{rgb}{0.0, 0.45, 0.0}
\definecolor{darkpurple}{rgb}{0.5, 0.0, 0.5}
\definecolor{darkbrown}{rgb}{0.4, 0.26, 0.13}

\newcommand{\CD}[1]{{\color{black}#1}}

\newcommand{\FJ}[1]{{\color{black}#1}} % changes by Frederic Jourdin

\newcommand{\eg}{\textit{e.g.,}\xspace}

\newcommand{\x}{\ensuremath{x}\xspace}
\newcommand{\y}{\ensuremath{y}\xspace}

\newcommand{\BBP}{\text{BBP443}\xspace}
\newcommand{\SPM}{\text{SPM}\xspace}
\newcommand{\CHL}{\text{Chl-a}\xspace}

\begin{document}

\title{Generalization performance of neural mapping schemes for the space-time interpolation of satellite-derived ocean colour datasets}%\\

% \author{
%     \IEEEauthorblockN{
%         Thi Thuy Nga Nguyen\IEEEauthorrefmark{1}\textsuperscript{§}
% ,
%         Clément Dorffer\IEEEauthorrefmark{1}\textsuperscript{§}
% ,
%         Frédéric Jourdin\IEEEauthorrefmark{2},
%         Ronan Fablet\IEEEauthorrefmark{1}
%     }
    
%     \IEEEauthorblockA{\IEEEauthorrefmark{1}IMT Atlantique, UMR CNRS Lab-STICC, INRIA team Odyssey, Brest, France}
    
%     \IEEEauthorblockA{\IEEEauthorrefmark{2}Service Hydrographique et Océanographique de la Marine (Shom), Brest, France}

\author{
    \IEEEauthorblockN{
        Thi Thuy Nga Nguyen
,
        Clément Dorffer
,
        Frédéric Jourdin,
        Ronan Fablet
    }

    \thanks{
Thi Thuy Nga Nguyen, Clément Dorffer and Ronan Fablet are with the IMT Atlantique, UMR CNRS Lab-STICC, INRIA team Odyssey, Brest, France.}

\thanks{
Frédéric Jourdin is with the Service Hydrographique et Océanographique de la Marine (Shom), Brest, France.}

    \thanks{
Thi Thuy Nga Nguyen and Clément Dorffer contributed equally to this work.}
\thanks{Clément Dorffer completed this work while at the IMT-Atlantique, UMR CNRS Lab-STICC, INRIA team Odyssey, Brest, France.}
    \thanks{This work was funded by the French ANR, as part of the IA-Biodiv Challenge: Fish-Predict; it was also supported by CNES (French Space Agency);  OceaniX (ANR-19-CHIA-0016) and EU Horizon Europe project EDITO Model Lab (Grant 101093293). It benefited from HPC and GPU resources from GENCI-IDRIS (Grant 2021-101030) and CPER AIDA GPU cluster supported by The Regional Council of Brittany and FEDER.}
}

\markboth{Journal of \LaTeX\ Class Files,~Vol.~14, No.~8, August~2021}%
{Shell \MakeLowercase{\textit{et al.}}: A Sample Article Using IEEEtran.cls for IEEE Journals}
%%%%% \IEEEpubid{2021 IEEE}

\maketitle

\begin{abstract}
Neural mapping schemes have become appealing approaches to deliver gap-free satellite-derived products for sea surface tracers. The generalization performance of these learning-based approaches naturally arises as a key challenge. This is particularly true for satellite-derived ocean colour products given the variety of bio-optical variables of interest, as well as the diversity of processes and scales involved. Considering region-specific and parameter-specific neural mapping schemes will result in substantial training costs. 
This study addresses generalization performance of neural mapping schemes to deliver gap-free satellite-derived ocean colour products. We develop a comprehensive experimental framework using real multi-sensor ocean colour datasets for two regions (the Mediterranean Sea and the North Sea) and a representative set of bio-optical parameters (Chlorophyll-a concentration, suspended particulate matter concentration, particulate backscattering coefficient). We consider several neural mapping schemes, and we report excellent generalization performance across regions and bio-optical parameters without any fine-tuning using appropriate dataset-specific normalization procedures. We discuss further how these results provide new insights towards the large-scale deployment of neural schemes for the processing of satellite-derived ocean colour datasets beyond case-study-specific demonstrations.
%Our findings demonstrate that this approach ensures robust generalization capabilities of neural mapping schemes across varied regions and variables. Importantly, this method significantly curtails the need for multiple specialized models, thereby reducing computational expenses while maintaining competitive performance with a single, versatile model. This advancement sets a new benchmark for cost-effective, scalable applications in satellite oceanography.
%, an innovative integration of deep neural networks with variational data assimilation proposed in \cite{Fablet2021Learning}.  Tackling the challenge of processing diverse  ocean color variables across different regions - where processing with multiple models can be resource-intensive - our work explores the generalization capacity of 4DVarNet that allows one model to fit all. Through extensive experimentation with real datasets from the North Sea and the Mediterranean Sea, we reveal the generalization performance of 4DVarNet in adapting and scaling across  diverse geographical domains and bio-optical parameters, even without any fine-tuning. Additionally, we provide insights into the intuitive reasons behind the effectiveness of the deep learning scheme in generalizing, which we believe adds notable value to the field of ocean color remote sensing in reducing training costs in wide range scenarios and large-scale applications.}
\end{abstract}

\begin{IEEEkeywords}
space-time interpolation; transfer learning; cross-modal learning; scalable algorithm; domain adaptation; data-driven model; data assimilation; image gap filling; suspended particulate matter; observing system experiment (OSE); ocean colour remote sensing; end-to-end deep learning; bio-optical parameter estimation; deep learning in satellite imagery.
\end{IEEEkeywords}
 
% \begin{keywords}
% interpolation; data-driven model; variational data assimilation; missing
% data; suspended particulate matter; observing system experiment; ocean colour remote sensing; Mediterranean; North Sea.
% \end{keywords}
%
\section{Introduction}
\label{sec:intro}

Satellite-derived ocean colour products provide invaluable data to monitor, model, and forecast bio-optical properties of the ocean \cite{Groom2019oceancolour}. Among others, they provide means to monitor the primary production of the ocean on a global scale \cite{westberry2023gross}, infer climate change \cite{dutkiewicz2019ocean,goes2020ecosystem}, model and forecast sediment dynamics in coastal areas \cite{Wei2021}, monitor Harmful Algal Blooms (HABs) and Aquaculture \cite{sathyendranath2014reports,davidson2016forecasting}, eventually assess biodiversity through Phytoplankton Functional Types \cite{alvain2005remote,xi2020global}, and various other applications of societal benefits \cite{platt2008ocean}.

We can distinguish three main categories of variables in ocean color remote sensing datasets: chlorophyll-a concentration, related to the phytoplankton which obeys the dynamics of primary production depending on the availability of light and nutrients \cite{FalkowskiRaven2007}; coloured dissolved organic matter, mostly related to refractory organic material, a by-product of the decomposition of organic material finally disappearing with sunlight bleaching \cite{coble2007marine}; and suspended particulate matter, a bulk concentration of the entire particulate assemblages, typically related to the living and detrital material in the open ocean and lithogenic material in the coastal ocean, the latter mainly depending on sediment load brought by rivers or resuspended from the bottom of the ocean \cite{Wei2021}. Other parameters of interest include particulate optical backscattering, which is known to evaluate the carbon content of organic particles and phytoplankton in the open ocean \cite{Bisson2020} and to observe mineral particles in the coastal ocean \cite{Boss2009}.

Satellite-derived ocean colour products involve several multispectral satellite sensors, such as SeaWiFS, MODIS-Aqua \& Terra, MERIS, VIIRS-SNPP \& JPSS1, OLCI-S3A \& S3B. These sensors differ in their sampling patterns and involve significant sampling gaps affected by cloud cover. The retrieval of gap-free products from such Multi-Sensor datasets then arise as a key challenge. Recently, neural mapping schemes \cite{Beauchamp2022, Fablet2021Learning, Barth2022, Wang2022} have emerged as attractive solutions to improve the quality of operational products, which generally smooth out fine-scale patterns. Current neural mapping demonstrations for ocean colour products \cite{YUAN2020_DLRemoteSensing} involve mainly region-specific and variable-specific case studies. As identified in \cite{DLChallenges}, speeding up computing time and reducing resource requirements are critical challenges, especially when scaling these approaches to multi-parameter configurations and global applications.

In this context, this study aims to assess the generalization performance of neural mapping schemes across ocean colour parameters and ocean regions. 
Among the state-of-the-art deep learning schemes for image gap filling in such complex datasets \cite{Fablet2021Learning, Barth2022, Wang2022}, we consider a classic UNet architecture \cite{UNet2015}, which is among the baseline computer vision architectures for a wide range of image-to-image mapping problems, and the 4DVarNet framework \cite{Fablet2021Learning}, which draws inspiration from variational data assimilation \cite{Bannister2017VariationalDA, Rabier2003VariationalDA} and optimal interpolation. The 4DVarNet framework has been previously shown to deliver state-of-the-art performance for the space-time interpolation of geophysical dynamics \cite{Beauchamp2022, Vient2022, DorfferEtAl}. 
Our main contribution is two-fold. We exploit the neural schemes (4DVarNet and UNet) with a learning strategy based on real gappy datasets and we design benchmarks accounting for:
\begin{itemize}
    \item three different variables: the bio-optical particulate backscattering coefficient (\BBP, in \unit{\per\meter}), the phytoplankton chlorophyll-\textit{a} concentration (\CHL, in \unit{\milli\gram\per\cubic\meter}%$mg \cdot m^{-3}$
 ) and the suspended particulate matter (\SPM, in \unit{\milli\gram\per\liter});
    \item and two areas: the Mediterranean Sea and the North Sea domain.
\end{itemize}
Our experiments showcase a remarkable generalization capacity
of the benchmarked neural schemes. They apply relevantly beyond the training datasets. These results further support the operational exploration of neural mapping schemes for ocean colour datasets, while keeping the computational requirements for training purposes moderate, including the potential re-use of pre-trained models for new variables and domains.

This paper is structured as follows. Section 2 presents the specifics of the chosen case studies, focusing on regions like the North Sea and the Mediterranean Sea, and highlights the bio-optical variables of the ocean color datasets utilized. Section 3 outlines the methodologies employed, especially the considered 4DVarNet scheme. In Section 4, we present and discuss our experiments and findings. Section 5 summarizes our main contributions and potential avenues for future research.

\section{Ocean colour datasets}
\label{sec:Case_study}

\begin{figure*}[ht]
    \centering
    % First subfigure
    \includegraphics[width=.55\textwidth]{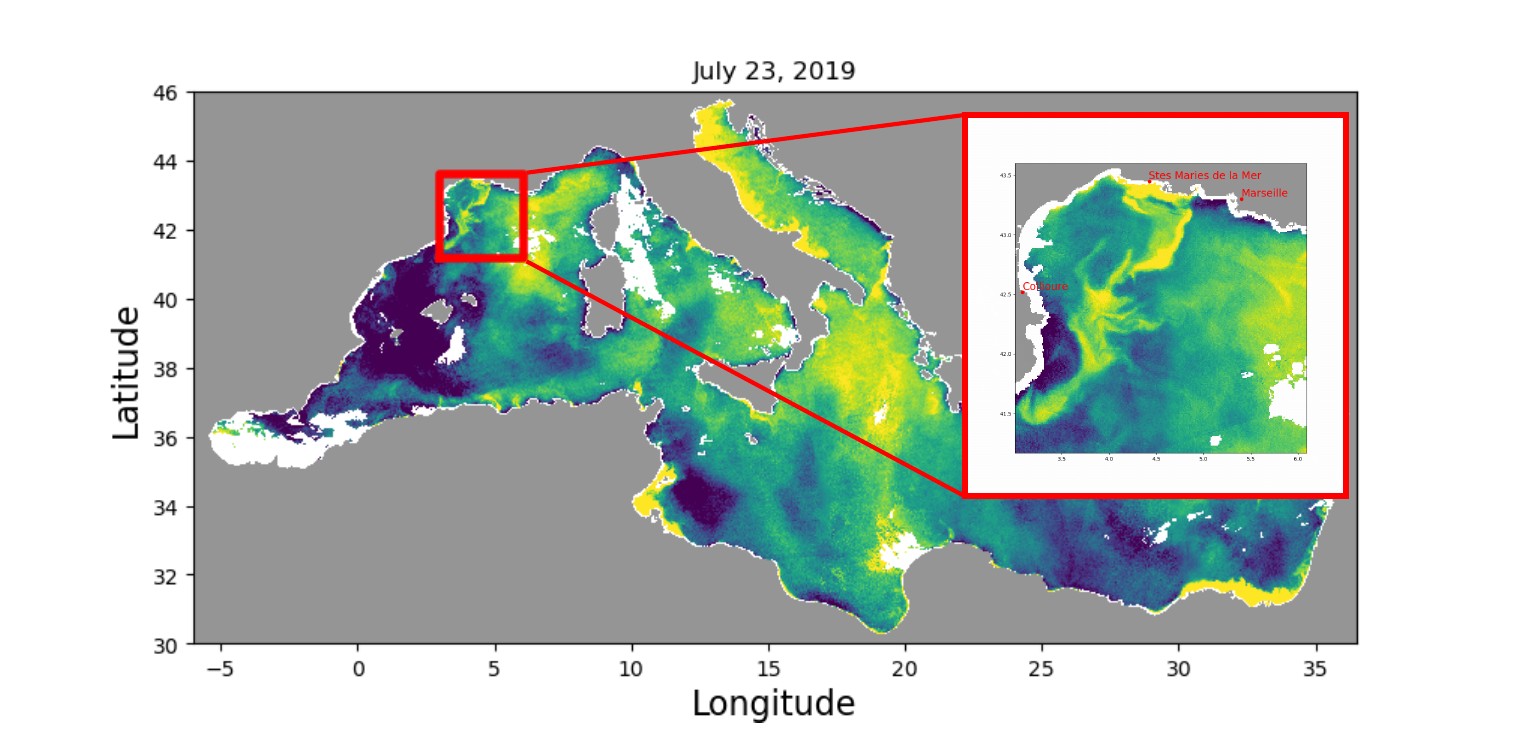} % Adjust the image path and width as necessary
    % Second subfigure
    \includegraphics[width=.35\textwidth]{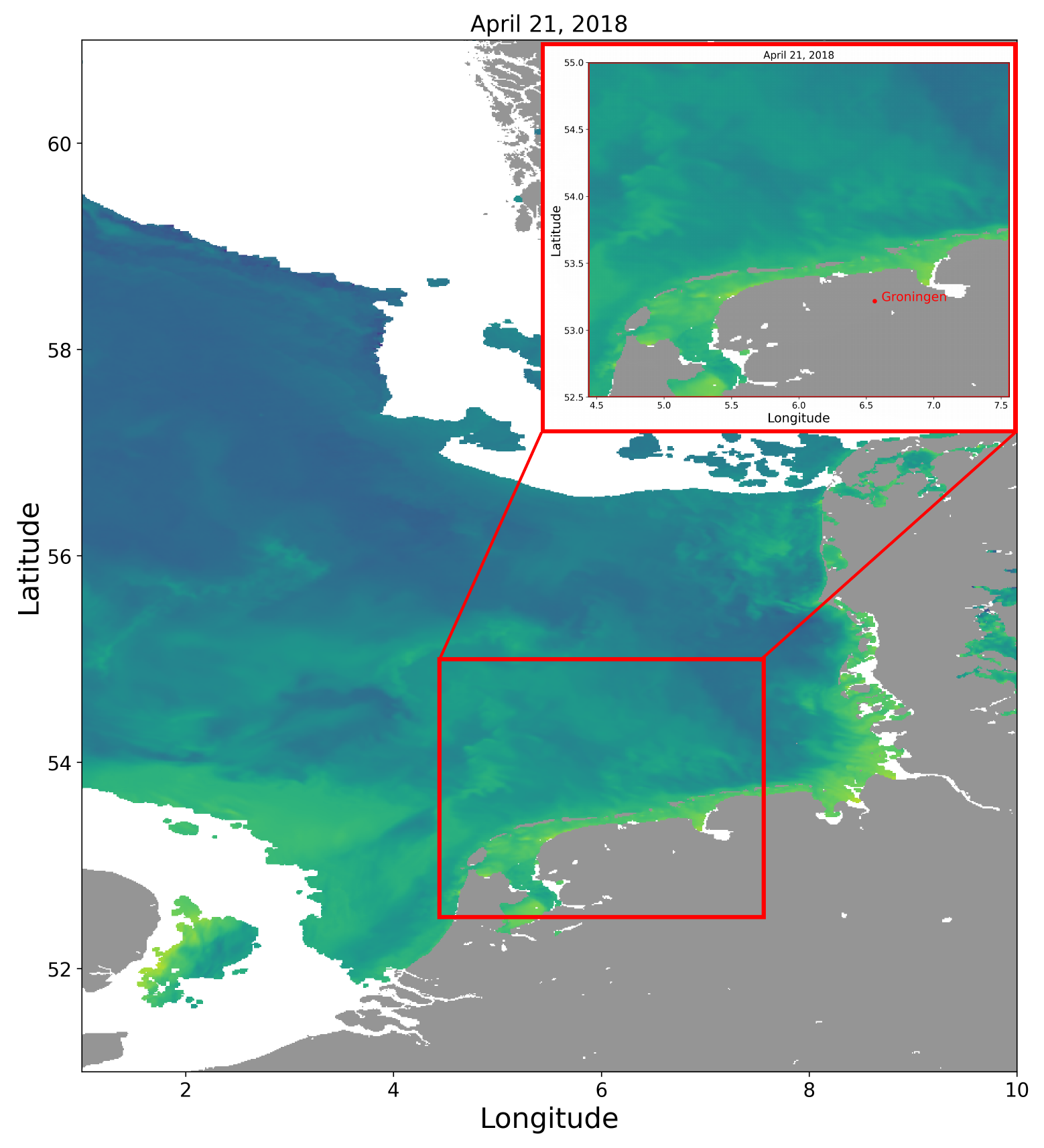}
    \caption{Areas and variables used as case studies. Left: \BBP in the Mediterranean Sea, selected area is the area used in training (Gulf of Lion). Right: \SPM concentration in the North Sea, selected area is Dutch Wadden Sea.}
    \label{fig:Selected_Areas}
\end{figure*}

% \begin{figure}[h]
%     \centering
%     % Subfigure a
%     \begin{subfigure}[]{0.45\textwidth} % Adjust the width to fit your document
%         \centering
%         \includegraphics[width=\textwidth]{./images/Merditeranean_area_selection.pdf} % Adjust the image path and width as necessary
%         \caption{\BBP in the Mediterranean Sea. Selected area is the area used in training (Gulf of Lion).}
%         \label{fig:Selected_Areas_sub1}
%     \end{subfigure}
%     \hfill % Adds some space between the subfigures
%     % Subfigure b
%     \begin{subfigure}[]{0.4\textwidth} % Adjust the width to fit your document
%         \centering
%         %\includegraphics[width=\textwidth]{./images/plot_4DVarNet_firstcaseStudy_2nd.pdf}
%         \includegraphics[width=\textwidth]{./images/plot_4DVarNet_NorthSea_report__April 21, 2018.pdf}
%         \caption{\SPM concentration in the North Sea. Selected area is Dutch Wadden Sea.}
%         \label{fig:Selected_Areas_sub2}
%     \end{subfigure}
    
%     \caption{Areas and variables used as case studies.}
%     \label{fig:Selected_Areas}
% \end{figure}

We use two datasets extracted from the Copernicus Marine Environment Monitoring Service (CMEMS), specifically the Level 3 (L3) biogeochemical layers products for the Mediterranean Sea \cite{cmems-mediterranean} and  for the North Sea \cite{cmems-northsea}, as shown in \cref{fig:Selected_Areas}. Our experimental design covers two coastal areas, a \si{240\kilo\meter}$\times$\si{300\kilo\meter} around the Dutch Wadden Sea and \si{240\kilo\meter}$\times$\si{240\kilo\meter} area in the Gulf of Lion, as well as the whole Mediterranean Sea. These three domains involve different ocean dynamics both for ocean physics, ocean sediments, and ocean primary production. As such, they offer us the means to assess the generalization performance across different dynamical regimes.

%For computational efficiency and to ensure representativeness of the data, we selected a smaller area which, despite its size, reasonably represents the entire larger region. Consequently, all proposed experiments are performed on restricted regions shown in \cref{fig:Selected_Areas}. For the Mediterranean Sea, we extract and consider a \si{240\kilo\meter}$\times$\si{240\kilo\meter} area picked in the South of France Mediterranean Sea, while for the North Sea, we extract and consider a \si{240\kilo\meter}$\times$\si{300\kilo\meter} area picked around the Dutch Wadden Sea. They both offer a \si{1\kilo\meter} resolution and daily measurements. These two regions involve different upper ocean and coastal dynamics.

We also emphasize that the considered ocean colour datasets involve different L2-to-L3 processings. This potentially leads to different noise patterns to assess the robustness of the mapping schemes to noise patterns, which are not present in the training dataset. We also point out that the missing data rates in the considered subregion of the North Sea are on average slightly higher than in the considered subregion of the Mediterranean Sea (60\% vs. 45\%).

%However, we show that the 4DVarNet algorithm adapts well across these differences, demonstrating that the algorithm is capable of generalizing to varied areas and variables. The demonstrations are latter shown in  \cref{sec:SameAreaVariable,sec:SameAreaDiffVariable,sec:SameVariableDiffArea,sec:DiffAreaVariable}. 

Moreover, when addressing reconstruction over an entire ocean basin, such as the entire Mediterranean Sea (Med. Sea), training on such a large area can suffer excessively high costs. To address this, we propose a strategy of training in a smaller area (selected area on the left in \cref{fig:Selected_Areas}) and then applying the model to the entire large region by segmenting it into multiple smaller sections. This approach aims at reducing training costs while maintaining comprehensive visualization, as shown in \cref{sec:SegmentedApplication}.

\section{Methods}
\label{sec:Methods}
This section introduces the neural mapping schemes under consideration: 4DVarNet \cite{Fablet2021Learning} and UNet \cite{UNet2015}. It also details our training configurations and the evaluation metrics used in our analysis.  

%We assess the generalization performance of the neural mapping schemes across various variables and domains without any fine-tuning. While both mapping schemes (4DVarNet and UNet) later demonstrate robust generalization capabilities, 4DVarNet particularly excels due to its sophisticated architecture that integrates a data assimilation framework, enhancing its ability to generalize effectively. Given these advantages, our focus will predominantly be on exploring and detailing the performance and applications of 4DVarNet. We will dedicate a substantial portion of this section to discussing the advanced features of 4DVarNet and will present more extensive experiments involving this model.

\subsection{4DVarNet scheme}

% \begin{figure*}[ht] % 'htbp' option specifies figure placement preference: here, top, bottom, separate page
%     \centering
%     \includegraphics[width=0.7\textwidth]{./images/4DVarNet_framework_AllRedraw_MedSea_example.drawio.png} % Adjust the image path and width as necessary
%     \caption{Sketch of the end-to-end deep learning architecture 4DVarNet.} 
%     \label{fig:4DVarNetFramework} 
% \end{figure*}

When using 4DVarNet schemes, the space-time interpolation relies on a variational formulation with a minimization problem defined as following:
\begin{align}
    \hat{x} &= \arg\min_{x} \mathcal{U}(x, y, \Omega) \nonumber \\
    &= \arg\min_{x} \left\{ \lambda_1 \left\| x - y \right\|^2_{\Omega} + \lambda_2 \left\| x - \phi(x) \right\|^2 \right\},
    \label{eq:variational_problem}
\end{align}
where \(\Omega\) denotes the observation domain, \(y\) represents the gappy observation data, and \(\phi\) is a defined prior. When applied to sea surface parameters, \(\phi\) is expected to encode the underlying space-time dynamics. Coefficients \(\lambda_1\) and \(\lambda_2\) balance the contribution of different terms in the variational cost function  \(\mathcal{U}\). More precisely, \(\lambda_1\) scales the data fidelity term, ensuring that the solution is consistent with the observations.  On other hands, \(\lambda_2\) adjusts the influence of the dynamical prior \(\phi(x)\), imposing smoothness or physical constraints on the solution. The choice of \(\lambda_1\) and \(\lambda_2\) determines the relative weight of observation fitting versus smoothness or model compliance.

In model-driven schemes \cite{Evensen2009DataAssimilation}, operator \(\phi\) involves a prior stated as an Ordinary or Partial Differential Equation. However, here, the considered 4DVarNet parameterization shifts towards neural priors where \(\phi\) exploits a Neural-Network-based architecture, such as \eg Convolutional Auto-encoder or U-Net to be trained from data.

Generally, problem (\ref{eq:variational_problem}) can be solved by gradient descent with successive updates:
\begin{equation*}
    x^{k+1} = x^k - \alpha \nabla_x\mathcal{U}(x^k, y, \Omega),
%\label{eq:gradient_descent}
\end{equation*}
where $x^{(k+1)}$ is the updated state, $x^{(k)}$ is the current state, and \(\alpha\) is a scalar parameter standing for a learning rate. However, optimization in the data-driven field often faces challenges such as including sparsity, noise, and imbalanced data points, non-convexities in very high-dimensional spaces, which may affect the optimization process where gradient descent might not be efficient. Inspired by the meta-optimizer concept of "learning to learn" gradient descent \cite{MetaOptimizer2016}, the 4DVarNet scheme unfolds  successive residual units to deliver an end-to-end architecture and provides a trainable iterative gradient-based  solver that minimizes the variational cost \(\mathcal{U}\). The associated iterative update rule is given by:
\begin{equation}
    x^{(k+1)} = x^{(k)} - T \left( LSTM \left[ \alpha \cdot \nabla_x \mathcal{U}_{\phi}(x^{(k)}, y, \Omega), h^{(k)}, c^{(k)} \right] \right)
    \label{eq:learnable_gradient_descent}
\end{equation}
where $T$ is a linear mapping, $LSTM$ is a unit of convolutional LSTM model \cite{ConvLSTM2015} that processes the gradients along with hidden state $h^{(k)}$ and cell state $c^{(k)}$, $\alpha$ is a scalar parameter, and $\nabla_x \mathcal{U}_{\phi}$ denotes the gradient of the cost function $\mathcal{U}_{\phi}$ with respect to the state $x$ at iteration $k$. We call \cref{eq:learnable_gradient_descent} the solver (of the variational minimization problem \cref{eq:variational_problem}). Overall, when a trainable neural prior $\phi$ is incorporated in 4DVarNet, the training phase concurrently learns both the prior and the solver, by optimizing the final reconstruction performance.

%Its whole architecture is shown in \cref{fig:4DVarNetFramework}. This scheme takes as input partial observations with a corresponding missing data mask and an initial guess for the unknown state to be reconstructed. It employs an iterative, gradient-based solver that utilizes a convolutional LSTM with a series of residual blocks. Each block takes the gradient of the variational cost with respect to the state \(x\), calculated from the output of the preceding residual step, as its input.  This allows 4DVarNet to iteratively refine its predictions. These gradients are efficiently computed using automatic differentiation tools available in PyTorch (with autograd function).}

\subsection{UNet scheme}
The UNet architecture \cite{UNet2015} is well-known for its versatility in image processing tasks. It offers many configuration choices, such as the number of layers, parameters, and skip connections. We empirically selected the best-performing configuration among the settings we experimented with. In this configuration, the model downsamples and upsamples once with a factor of 2, with a hidden dimension size of 128, resulting in a total of 625 thousand parameters. The bottleneck uses bilinear residual layer which is effective in capturing intrinsic non-linearities of physical dynamical systems as shown in \cite{Fablet2017Bilinear}.

\subsection{Data processing and data normalization}

\begin{figure*}[ht]
    \centering
    % BBP Observation
    \includegraphics[width=.22\linewidth]{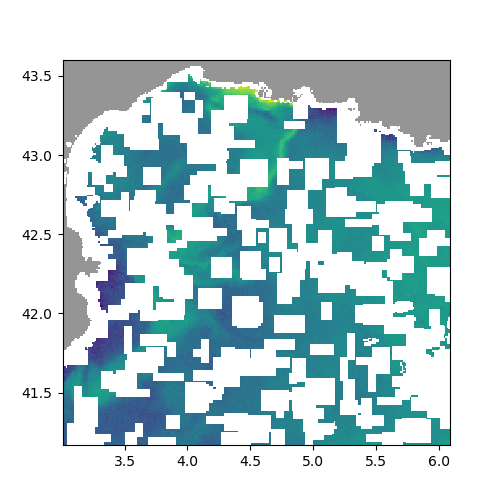} % Adjust the image path and width as necessary
    % BBP Satellite Image
    \includegraphics[width=.22\linewidth]{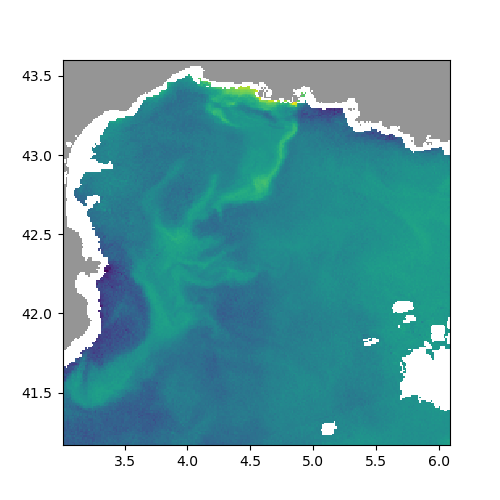}
    % SPM Observation
    \includegraphics[width=.25\linewidth]{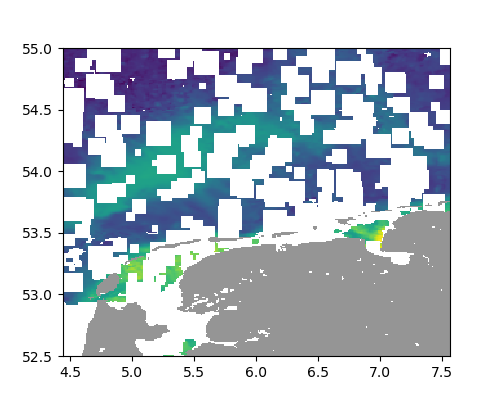}
    % SPM Satellite Image
    \includegraphics[width=.25\linewidth]{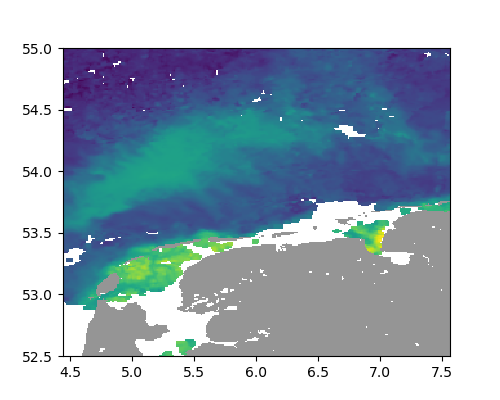}
    
    \caption{Examples of simulated observations. Left to right: BBP simulated observation and satellite image for the Gulf of Lion area (the restricted area in the Med. Sea), SPM simulated observation and satellite image for the Wadden Sea area.}
    \label{fig:OnlineObservations}
\end{figure*}

For data processing, we prepare two sets of images: one for observation \y and another for the target to compare with the reconstruction \x.  The observation set comprises images with artificially removed patches to mimic cloud cover effects (shown in Figures in \cref{fig:OnlineObservations}). During the training phase, the reconstructed output is compared against the corresponding satellite image, which serves as the target, using the mean square error (MSE) loss. This loss is computed only for pixels that are available (visible) in the target.

% \begin{figure*}
%     \centering
%     \begin{subfigure}{0.48\linewidth}
%         \centering
%         \includegraphics[width=\textwidth]{./images/Obs_BBP.png}
%         \caption{Observation \(y\).}
%         \label{fig:OnlineObservations_Random_ObservationBBP}
%     \end{subfigure}
%     \begin{subfigure}{0.48\linewidth}
%         \centering
%         \includegraphics[width=\textwidth]{./images/target_BBP.png}
%         \caption{Satellite Image/ Target.}
%         \label{fig:OnlineObservations_SatelliteImageBBP}
%     \end{subfigure}
    
%     \begin{subfigure}{0.48\linewidth}
%         \centering
%         \includegraphics[width=\textwidth]{./images/Obs_SPM.png}
%         \caption{Observation \(y\).}
%         \label{fig:OnlineObservations_Random_ObservationSPM}
%     \end{subfigure}
%     \begin{subfigure}{0.48\linewidth}
%         \centering
%         \includegraphics[width=\textwidth]{./images/target_SPM.png}
%         \caption{Satellite Image/ Target.}
%         \label{fig:OnlineObservations_SatelliteImageSPM}
%     \end{subfigure}
    
%     \caption{Online simulated observation examples for \SPM of (1st line) Med. Sea area and (2nd line) North Sea area.}
%     % \caption{Data and online simulated observation example.}
%     \label{fig:OnlineObservations}
% \end{figure*}

 Here, the random observations (with randomly removed patches) are generated online, meaning that they are produced on-the-fly  during the training process. This approach helps us enrich the dataset by providing more training samples and can be considered as a data augmentation method \cite{DorfferEtAl}.

 In our approach, we normalize each dataset to standardize their distributions, which is crucial for both the training and inference phases. For each dataset, we first calculate its mean \( m \) and variance \( \sigma \). Note that each dataset has its own and different mean \( m \) and variance \( \sigma \). We then apply 
normalization by subtracting the mean from each data point and dividing by the standard deviation. As described below: 
\begin{equation*}
x_{\text{normalized}} = \frac{x - m}{\sqrt{\sigma}},
\end{equation*}
where  \( x \) is a data point in the dataset.
This process aligns the datasets to a common scale, enhancing the model's ability to transfer learning from one dataset to another.

\subsection{The transfer workflow}
\begin{figure*} % 'htbp' option specifies figure placement preference: here, top, bottom, separate page
    \centering
    \includegraphics[width=0.9\textwidth]{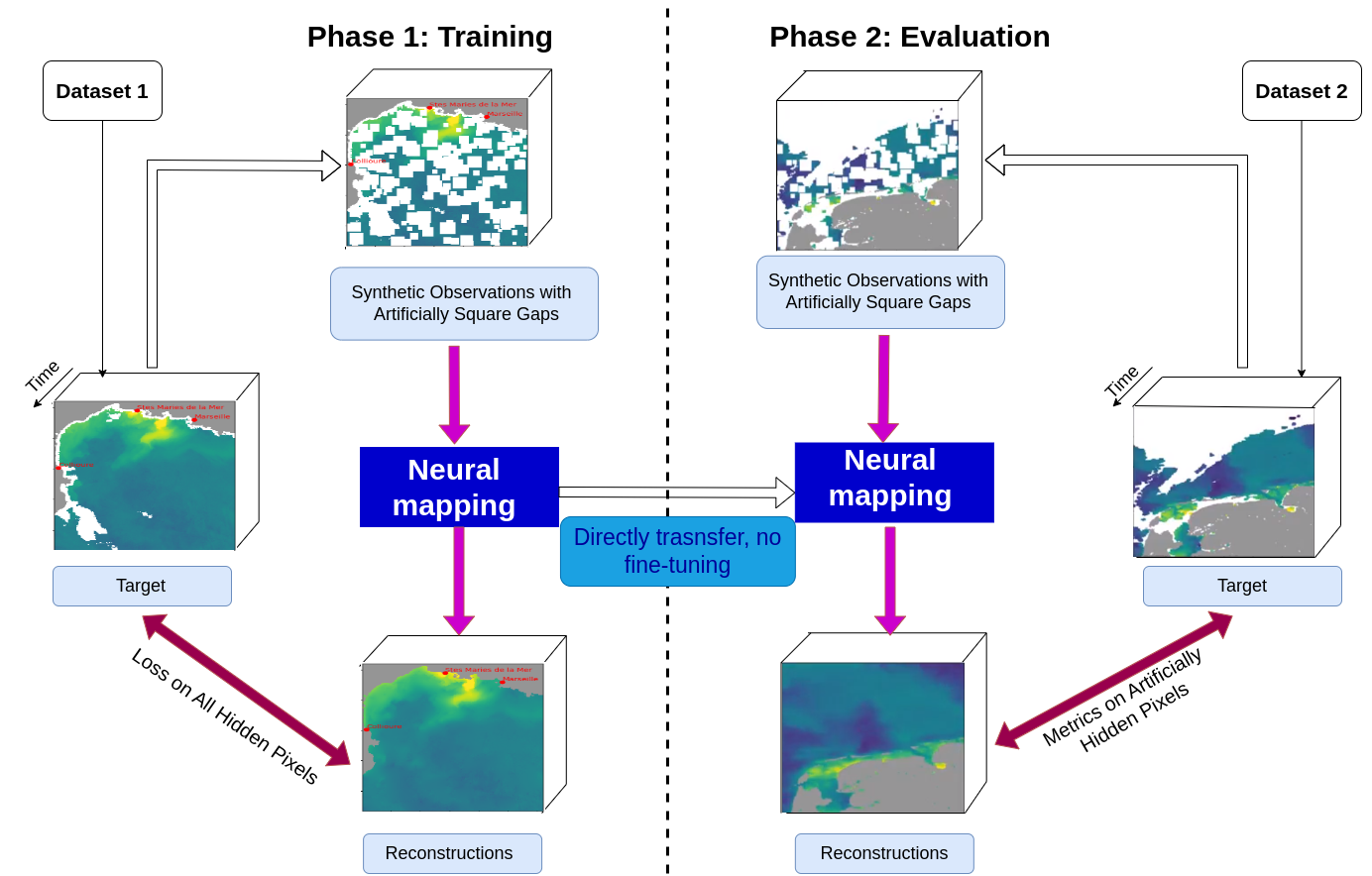} % Adjust the image path and width as necessary
    \caption{The transfer workflow for neural mapping. The training phase (left) involves training the model on one chosen dataset with synthetic observations that include artificially-generated gaps, aiming to learn gap-filling techniques. The evaluation phase (right) tests the model's generalization ability on another dataset, without further tuning, to assess generalization performance across new, unseen data.} 
    \label{fig:transferworkflow} 
\end{figure*}

The workflow is illustrated in \cref{fig:transferworkflow}, including the training phase of a baseline model on a chosen dataset and then the evaluation phase on other datasets to demonstrate the generalization capacity of the neural schemes. Below are detailed explanations:
\begin{itemize}
    \item \textbf{Training Phase:} The neural schemes are first trained on a chosen dataset. During this phase, the model learns the context of data gap filling, enhanced by supervised learning techniques. Particularly in the 4DVarNet scheme, thanks to the integration of variational cost functions within the neural networks, the model not only learns how to reconstruct images but also understands the underlying space-time dynamics.
    
    \item \textbf{Evaluation Phase:} The trained model is then applied directly to a different dataset, without any fine-tuning. This phase evaluates the model's capacity to generalize across datasets.
\end{itemize}

\subsection{Training and Evaluation Framework}
\label{subsec:training_evaluation}

For evaluation purposes, we trained four distinct neural mapping schemes using 4DVarNet and UNet architectures. The first model, designated as 4DVarNet-Med-BBP (or UNet-Med-BBP for the UNet version), serves as the baseline and is trained to interpolate \BBP (the BBP at the specific wavelength of \si{443\nano\meter}) data over the Mediterranean Sea area. It will be compared with the other three models as detailed in the following. The second model, 4DVarNet-Med-CHL (or UNet-Med-CHL), interpolates \CHL data in the same Mediterranean Sea area. The third model, designed for the North Sea, interpolates \BBP data and is referred to as 4DVarNet-North-BBP (or UNet-North-BBP). The fourth model, 4DVarNet-North-SPM (or UNet-North-SPM), interpolates \SPM data in the North Sea.  The fourth model 4DVarNet-North-SPM (or UNet-North-SPM) differs from the baseline 4DVarNet-Med-BBP (UNet-Med-BBP), both in terms of  the considered variable and of the targeted geographical area.
%Despite these differences, we will demonstrate the impressive generalization capabilities of the neural mapping schemes.

The training, validation, and testing periods for these models are shown in \cref{tab:model_periods}.

\begin{table}[ht]
\centering
\resizebox{0.5\textwidth}{!}{%
\begin{tabular}{|l|l|l|}
\hline
                    & \textbf{Med. Sea Dataset} & \textbf{North Sea Dataset} \\ \hline
\textbf{Variables}   & \BBP, \CHL                   & \BBP, \SPM              \\ \hline
\textbf{Training Period}    & Jan 1, 2017 - Dec 31, 2018  & Jan 1, 2015 - Dec 31, 2016 \\ \hline
\textbf{Validation Period}  & Jan 1, 2021 - Dec 31, 2021  & Jan 1, 2017 - Dec 31, 2017 \\ \hline
\textbf{Testing Period}     & Jan 1, 2019 - Dec 31, 2020  & Jan 1, 2018 - Dec 31, 2018 \\ \hline
\end{tabular}%
}
\caption{Training, validation, and testing periods of the 4DVarNet models.}
\label{tab:model_periods}
\end{table}

\subsection{Performance Metrics}
\label{sec:performance_metrics}

Regarding evaluation metrics, we employ two metrics: the Root Mean Square Log Error (RMSLE) and the Relative Error (RE) as detailed below. RMSLE, a modification of Root Mean Square Error with logarithmic scaling, is particularly suited for comparing predicted and observed strictly positive values where the predictions are expected to vary across several orders of magnitude. It is defined as:
\begin{equation*}
    RMSLE = \sqrt{\frac{1}{n} \sum_{i=1}^{n} \left( \log_{10}(P_{i}^{Pred}) - \log_{10}(P_{i}^{Obs}) \right)^2}
    \label{eq:rmsle}
\end{equation*}
where:
\begin{itemize}
    \item $P_{i}^{Obs}$, $P_{i}^{Pred}$ are observation and prediction of pixel
i-th, respectively.
    \item $n$ is the number of considered pixels.
    \item The logarithm base 10 emphasizes the validation of low concentrations.
\end{itemize}

The RE metric assesses the relative error compared to the true value. It is especially useful when comparing the accuracy of predictions across different scales or magnitudes and defined as:
\begin{equation*}
    RE = \frac{1}{n} \sum_{i=1}^{n} \left(\frac{|P_{i}^{Obs} - P_{i}^{Pred}|}{P_{i}^{Obs}} \times 100\%\right)
    \label{eq:relative_error}
\end{equation*}
This metric is expressed as a percentage, providing a clear understanding of the signal-to-error ratio, making it easier to interpret across different contexts and scales. Note that RMSLE and RE metrics are computed only on pixels that are available (visible) in the target but not in the observation.

\subsection{Reference Method for Comparison}

For benchmarking purposes, we select DInEOF \cite{DInEOFMethod} which is a state-of-the-art approach in ocean remote sensing that is used for the operational production of L4 satellite images\cite{Volpe_2018}. DInEOF is a data-driven technique that applies empirical orthogonal functions (EOFs) for gap filling and feature interpolation in oceanographic datasets.  

\section{Experiments}
\label{sec:Experiments}

In this section, we assess the generalization performance of the neural mapping schemes (4DVarNet and U-Net) across different variables and domains, without any fine-tuning. As mentioned earlier in Section \cref{subsec:training_evaluation}, we use the models trained on a restricted area of the Mediterranean Sea (the Gulf of Lion) with BBP data  as our baselines, namely the 4DVarNet-Med-BBP (also called 4DVarNet-Baseline) and UNet-Med-BBP (also called UNet-Baseline). We first report their performance for the associated test dataset (Section \ref{sec:SameAreaVariable}). We then assess their generalization when transferred to other variables and/or domains in the following sections. % \ref{sec:SameAreaDiffVariable}. 

\subsection{Baseline Performance for Med. Sea Area and \BBP variable}
\label{sec:SameAreaVariable}

\begin{table}[h]
    \centering
    \begin{tabular}{|c|c|c|}
        \hline
        Algo. & RMSLE & RE (\%)\\
       \hline
       DInEOF & 0.080 & 13.66\\
       % \hline
       UNet-Med-BBP (UNet-Baseline) & 0.075 & 10.99\\
       % \hline
       4DVarNet-Med-BBP (4DVarNet-Baseline) & \textbf{0.050} & \textbf{7.30}\\ 
       \hline
    \end{tabular}
    \caption{\BBP reconstruction performances obtained on the restricted Med. Sea area (Gulf of Lion).}
    \label{tabBBP_Med}
\end{table}

\begin{figure*}[ht]
    \centering
    \includegraphics[width=0.9\textwidth]
    {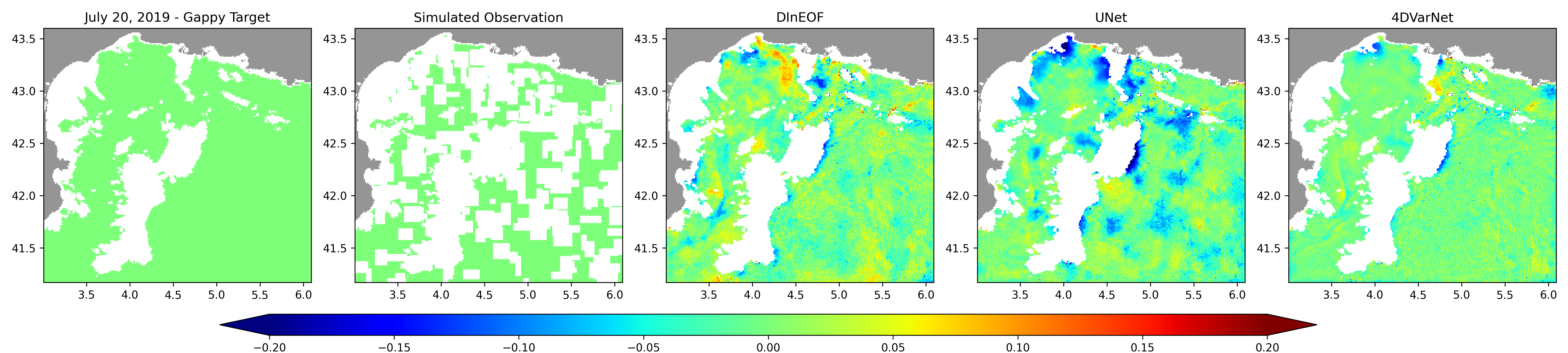}
    \caption{Comparative error maps of interpolated BBP443 fields in the Med. Sea on July 20, 2019. From left to right: Satellite Observation (Gap-Free Target) as reference, Simulated Observation with artificial gaps, and error maps for interpolated outputs from DInEOF, UNet, and 4DVarNet. The error scales are depicted in a log\(_{10}\) scale of $m^{-1}$, illustrating the interpolation accuracy of each method’s performance.}
    %\caption{Error maps of interpolated \BBP fields in the Mediterranean Sea area, depicted in log\(_{10}\) scale of $m^{-1}$.}
    \label{fig:error_map}
\end{figure*}

Table~\ref{tabBBP_Med} compares the performance of the 4DVarNet-Med-BBP and UNet-Med-BBP (Baseline models) trained and evaluated in the Gulf of Lion (Mediterranean Sea) with BBP dataset. Firstly, the two neural mapping approaches (4DVarNet and UNet) outperform DInEOF. 4DVarNet leads to the best performance, reporting a significant relative improvement of about 38\% in terms of RMSLE compared with DInEOF interpolation, and about 33\% compared with UNet interpolation.
This is further emphasized by the visual analysis of the interpolated fields displayed in Fig.\ref{fig:error_map}.  We can clearly observe that 4DVarNet model demonstrates the lowest error compared to UNet and DInEOF. Besides, the error map shows fewer and smaller regions of high error, indicating better overall accuracy and smoother interpolations. 

\subsection{Generalisation performance when Transferred to (\CHL)}
\label{sec:SameAreaDiffVariable}
In this section, we assess the performance of the 4DVarNet and UNet baselines, initially trained with \BBP, when applied to a different variable, \CHL, to illustrate their generalization from one variable to another one. For evaluation purposes, we also asses the 4DVarNet-Med-CHL and UNet-Med-CHL models which are trained using \CHL data in the Med. Sea area.

The results are presented in \cref{tabCHL_Med}. We highlight several key observations. First, both 4DVarNet-Baseline and UNet-Baseline demonstrate a good generalization capability, exhibiting only minor reductions in performance compared to models trained directly on the \CHL dataset and showing clear improvements over DInEOF, even when trained on a different variable. Interestingly, the two 4DVarNet schemes still significantly outperforms the other 
approaches. We observe only a negligible reduction in performance for 4DVarNet-Baseline compared to 4DVarNet-Med-CHL, despite the former being applied to a different variable than it was originally trained on. By contrast, the difference in the performance of the two UNet schemes is larger. This suggests a greater generalization of the 4DVarNet scheme.

Based on these results supporting a significantly better performance and generalisation of 4DVarNet schemes over UNet ones, we focus hereafter on 4DVarNet schemes with a view to assessing further the generalisation behaviour of neural mapping schemes across variables and regions.

\begin{table}[h]
    \centering
    \begin{tabular}{|c|c|c|}
        \hline
        Algo. & RMSLE & RE (\%)\\
       \hline
       DInEOF & 0.112 & 17.16\\
       UNet-Med-CHL  & 0.095 & 14.07 \\ 
       UNet-Baseline & 0.104 & 16.97 \\
       4DVarNet-Med-CHL & \textbf{0.058} & \textbf{8.15} \\ 
       4DVarNet-Baseline & 0.061 & 8.50 \\ 
       \hline
    \end{tabular}
    \caption{\CHL reconstruction performances obtained on the restricted Med. Sea area (Gulf of Lion).}
    \label{tabCHL_Med}
\end{table}

\subsection{Generalisation Performance when Transferred to North Sea Area}
\label{sec:SameVariableDiffArea}

In this section, we assess the performance of the 4DVarNet-Baseline model, initially trained with \BBP in the Med. Sea, when transferred to the North Sea using the same \BBP variable. To apply the 4DVarNet baseline trained on a 240x240 pixel region of the Med. Sea to a larger 240x300 pixel region of the North Sea, we divided the North Sea region into two overlapping 240x240 regions. After applying the model to each patch, we combined the results into a single reconstruction by averaging the values in the overlapping area. As reported in \cref{tabBBP_North}, the 4DVarNet-Baseline model is only marginally less accurate than 4DVarNet-North-BBP, which was specifically trained for the North Sea area. Notably, it still largely outperforms the DInEOF scheme.

\begin{table}[h]
    \centering
    \begin{tabular}{|c|c|c|}
        \hline
        Algo. & RMSLE & RE (\%)\\
       \hline
       DInEOF & 0.17 & 30.9 \\ 
       4DVarNet-North-BBP  & \textbf{0.12} & \textbf{18.35} \\ 
       4DVarNet-Baseline & 0.13 & 22.6  \\ 
       
       \hline
    \end{tabular}
    \caption{\BBP reconstruction performances obtained on the restricted North Sea area (Wadden Sea).}
    \label{tabBBP_North}
\end{table}

\subsection{Generalisation Performance when Transferred to \SPM in the North Sea Area)}
\label{sec:DiffAreaVariable}
In this section, we assess the 4DVarNet-Baseline model's performance, initially trained on the \BBP variable in the Med. Sea, when transferred to a geographically and geophysically different area - the North Sea - using a different variable, \SPM. This experiment aims to highlight the model's capability for both spatial and variable transfer. The results, presented in \cref{tabSPM}, indicate only a marginal decrease (below 5\%) in the interpolation performance of the 4DVarNet-baseline compared to the 4DVarNet-North-SPM, which is specifically trained for the North Sea area with \SPM. And the interpolation performance still remains highly competitive compared to the DInEOF scheme.

\begin{table}[h]
    \centering
    \normalsize % Adjust the font size, or \small 
    \setlength{\tabcolsep}{5pt} % Adjust the padding between table columns
    \begin{tabular}{|c|c|c|}
        \hline
        Algo. & RMSLE & RE (\%) \\
        \hline
        DInEOF & 0.181 & 33.35 \\
        4DVarNet-North-SPM & \textbf{0.127} & \textbf{18.77} \\
        4DVarNet-Baseline & 0.130 & 18.93 \\
        % 4DVarNet (trained on \CHL) & 0.129 & 19.40 \\  I comment this to have a consistence with other tables.
        \hline
    \end{tabular}
    \caption{\SPM reconstruction performances obtained on the restricted North Sea area (Wadden Sea).}
    \label{tabSPM}
\end{table}

\subsection{Generalisation to \BBP for the whole Mediteranean Sea}
\label{sec:SegmentedApplication}

% \begin{figure}[h]
%     \centering
% \includegraphics[width=1.0\linewidth]{images/08_22.png}
%     \caption{Full interpolation over the Med. Sea.}
%     \label{fig:full_interp}
% \end{figure}

We benefit from the fully-convolutional nature of the 4DVarNet architecture to apply it to the entire Med. Sea and report results for the mapping of \BBP.  To account for memory constraints, we adopt a patch-level approach. We split the whole domain into \CD{297 patches of dimension $240\times240$ that overlap by 106 pixels in latitude and 122 pixels in longitude. The interpolation is then performed independently for each patch prior to apply a post-processing procedure that delivers a gap-free reconstruction for the whole Med. Sea by merging all the interpolated patches.} %The later addresses patch boundary and overlap issues. This prevents the generation of patch-level artifacts.% as shown in \cref{fig:full_interp}.

% \begin{figure}[h]
%     \centering
% \includegraphics[width=1.0\linewidth]{images/Annual_Mean_Values_2019.png}
%     \caption{Mean value map over the Med. Sea of the whole year 2019.}
%     \label{fig:Annual_Mean_Values_2019}
% \end{figure}

\begin{figure*}[!t]
    \centering
    \subfloat[Monthly mean values for January 2019.]{%
        \includegraphics[width=.4\linewidth]{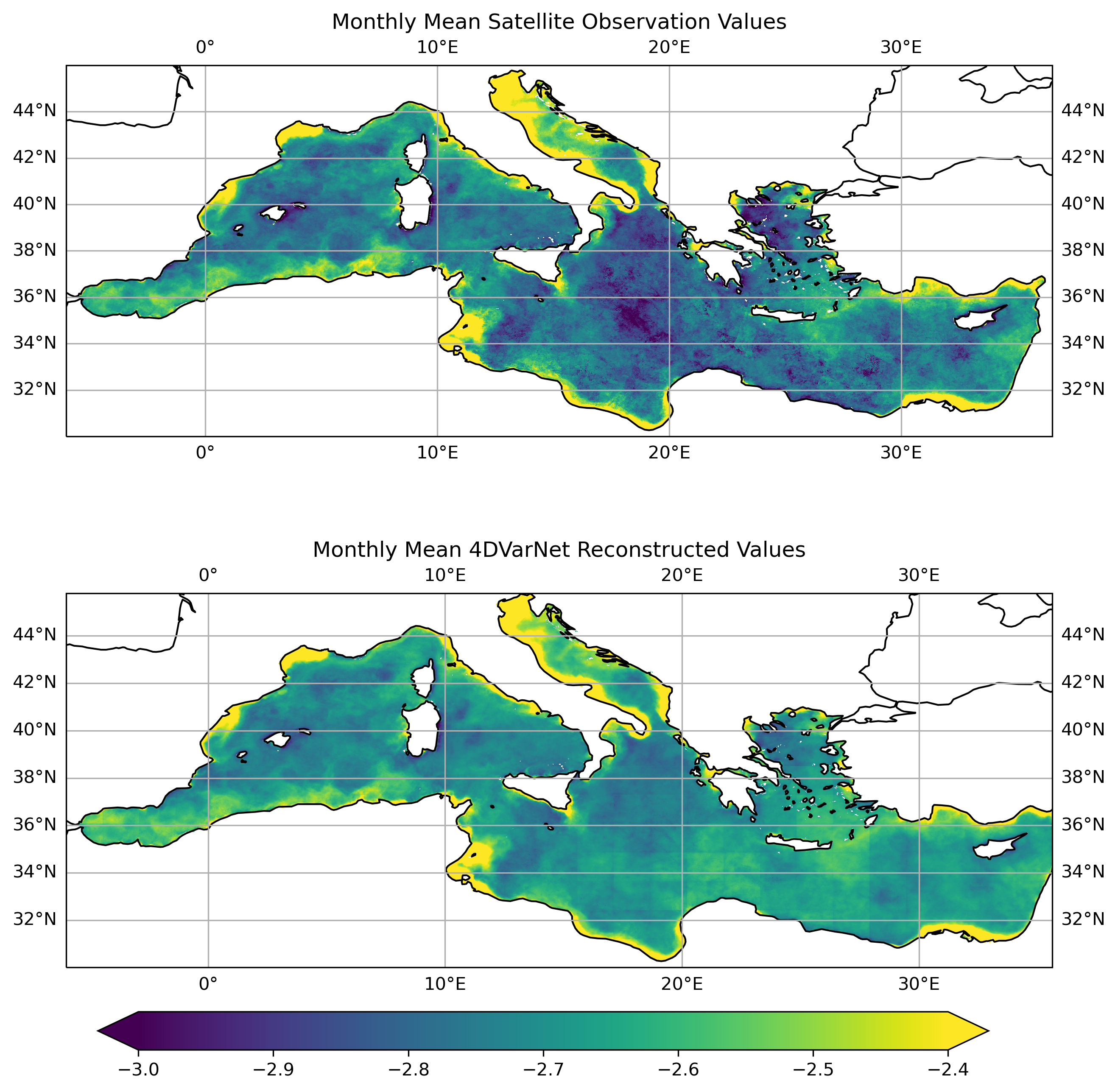}
        \label{fig:january}
    }
    \subfloat[Monthly mean values for July 2019.]{%
        \includegraphics[width=.4\linewidth]{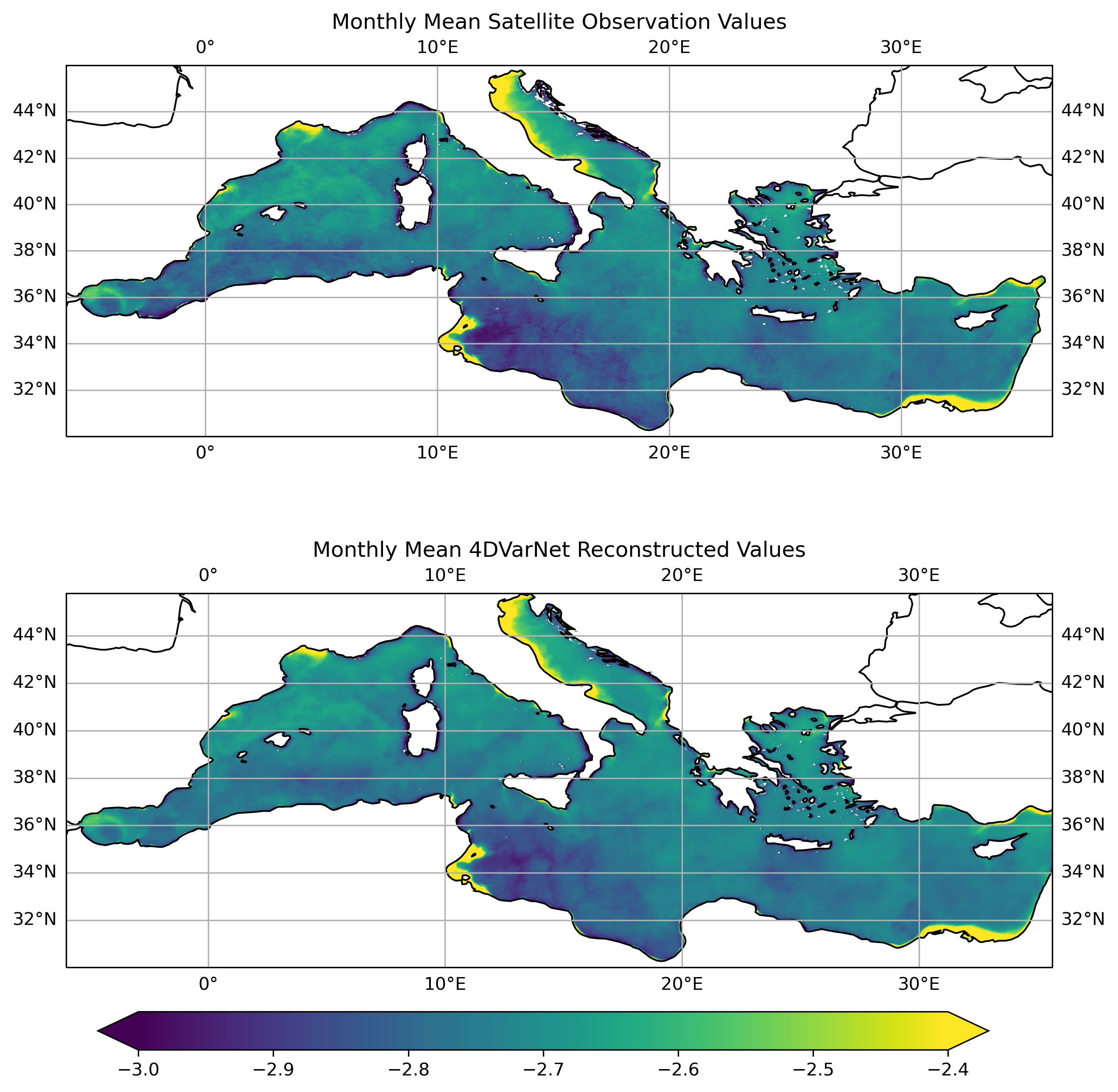}
        \label{fig:july}
    }
    %\caption{Comparative display of monthly mean values of the whole Med. Sea, depicted in log\(_{10}\) scale of $m^{-1}$.}
    \caption{Comparative display of monthly mean values for the Mediterranean Sea, January and July 2019, depicted in log\(_{10}\) scale of $m^{-1}$. Top panels: monthly mean of satellite observations; bottom panels: monthly mean of 4DVarNet reconstructions.}
    \label{fig:MonthlyMeans_wholeMedSea}
\end{figure*}

To asses the relevance of the gap-free fields over the entire Mediterranean Sea,
 we compare in \cref{fig:MonthlyMeans_wholeMedSea} monthly mean maps derived from satellite observations with those from 4DVarNet reconstructions for a selected month in summer (July) and another selected month in winter (January) of the year 2019. As expected, the irregular space-time sampling due to cloud cover affects the computation of monthly mean fields as exhibited by the patchy patterns we observe for the processing of the raw gappy satellite observations for January 2019. In contrast, 4DVarNet gap-free reconstruction produces a more comprehensive and smooth mean pattern, ensuring that mean values are consistent and not biased by the sampling pattern. Hence, its mean is more stable and reliable. We can notice that the differences are less noticeable for July 2019 as the cloud cover is much lower during the summer. Comparison of mean performance for January and July 2019 suggests that the effect of sampling may be even greater than that of seasonal variability.

\section{Discussion}

In this section, we discuss why neural mapping schemes demonstrate generalization performances on different ocean colour datasets, especially 4DVarNet with a higher generalization potential. A critical aspect of this finding can be attributed to the diversity of the baseline area used for training. This region encompasses both offshore and coastal areas, characterized by distinctly different water colours - from clear to turbid. Training the model across these different conditions likely enables it to learn and adapt to a broader range of scenarios, as opposed to training exclusively on homogeneous data (e.g., only clear open sea waters). This varied training environment challenges the model with complex patterns, potentially enhancing its ability to generalize to new, diverse conditions. In ocean color science, it is common to classify waters in two types: Case-1 and Case-2 waters \cite{morel1977analysis}. In Case-1 waters, all bio-optical parameter values recorded at the same time and at the same location mainly depend on a unique variable: Chla. By contrast, Case-2 waters are significantly influenced by other constituents such as lithogenic particles and dissolved material typically brought by rivers discharge or resuspended from the floor, typically found in some coastal areas. Case-2 waters are optically more complex and do not display the strong correlations stated before. The areas selected for training the baseline models (i.e, the continental shelf waters of the northern part of the Gulf of Lion and the North Sea) comprise both Case-1 and Case-2 waters \cite{lee2006global}. Overall, this diversity allows our baseline models to learn more complex patterns, thereby enhancing its ability to generalize when applied to other areas.

%In our areas of study, all continental shelf waters (the northern part of the Gulf of Lion and the whole North Sea) are typically Case-2 waters \cite{lee2006global}. Then those waters cannot display the correlations that would have explained the generalization performance that is observed in our present experiments with 4DVarNet.}

Another likely key aspect to consider is our data normalization scheme. By standardizing the distribution of each dataset, we ensure that the patterns are more consistent across different datasets after normalization. This consistency is likely a key factor in enabling the model trained on one dataset to perform effectively on others.

\begin{figure*}[!t]
    \centering
    \subfloat[Satellite Observations.]{%
        \includegraphics[width=.8\linewidth]{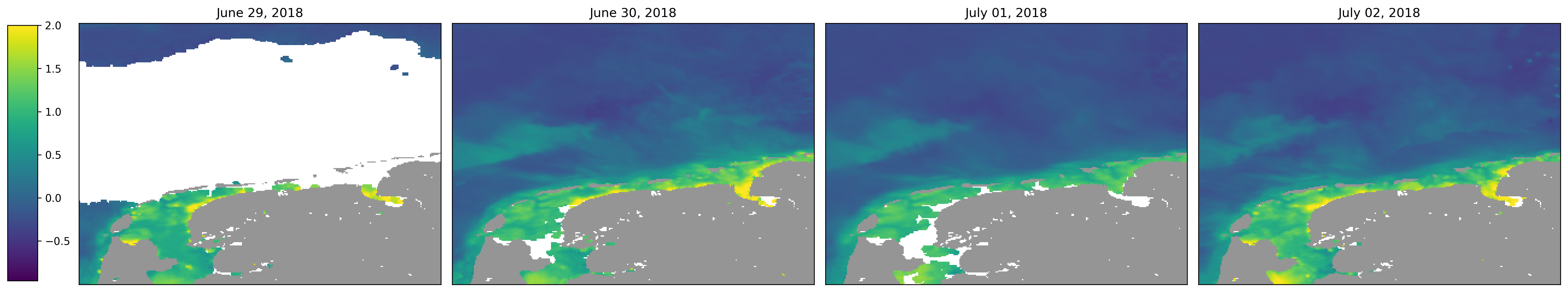}
        \label{fig:gradual_change_sub1}
    }
    
    \subfloat[4DVarNet reconstructions.]{%
        \includegraphics[width=.8\linewidth]{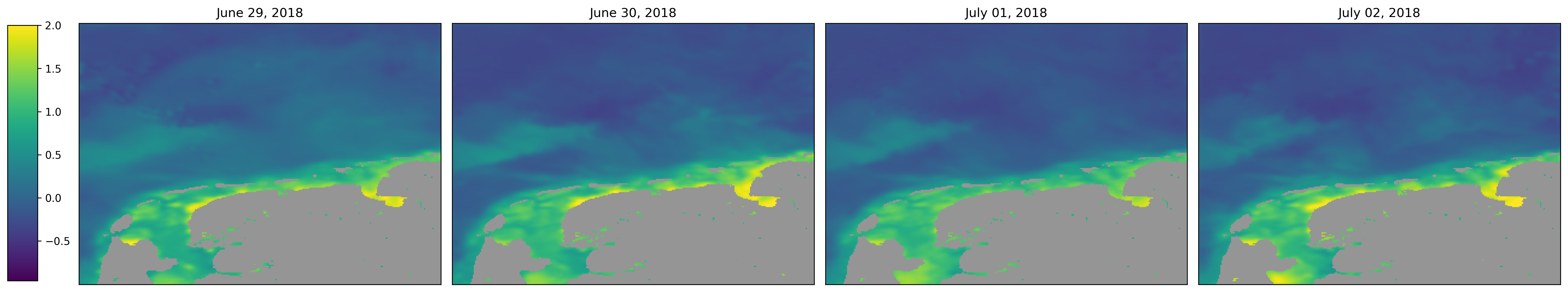}
        \label{fig:gradual_change_sub2}
    }
    \caption{Gradual change of \SPM in the Dutch Wadden Sea, shown in log\(_{10}\) scale of mg/l. Top panels depict consecutive days of gappy satellite observations; bottom panels show consecutive days of 4DVarNet reconstructions. Note: The datasets used do not ideally represent shallow waters near the coast, as discussed further in the Discussion section.}

    \label{fig:gradual_change}
\end{figure*}

Moreover, a noteworthy observation is the similarity of the inherent characteristics of different bio-optical variables, such as \BBP, \CHL, and \SPM. These variables often display similar changes, largely influenced by biological and chemical processes in the water, like phytoplankton growth cycles evolving over days to weeks. This results in a relatively smooth data profile (figure \ref{fig:gradual_change}), which might be different from physical parameters such as SSH and SST. Consequently, the similar variability of bio-optical variables seems to favour the generalization of interpolation performance across various bio-optical conditions, making the transferability achievable without additional learning steps, such as fine-tuning.

%SST can undergo rapid changes due to atmospheric conditions, such as heating by the sun or cooling by wind and evaporation, while SSH is subject to variations from tides, currents, and other physical oceanographic factors.  
% \begin{figure}
%     \centering
%     \begin{subfigure}{.5\textwidth}
%         \centering
%         \includegraphics[scale=0.16]{images/gradual_change_GT_DutchWaddenSea.png}
%         \caption{Satellite Observations.}
%         \label{fig:gradual_change_sub1}
%     \end{subfigure}%
    
%     \begin{subfigure}{.5\textwidth}
%         \centering
%         \includegraphics[scale=0.16]{images/gradual_change_DutchWaddenSea.png}
%         \caption{4DVarNet reconstructions.}
%         \label{fig:gradual_change_sub2}
%     \end{subfigure}
%     \caption{Gradual change of \SPM around the Dutch Wadden Sea.}
%     \label{fig:gradual_change}
% \end{figure}

Futhermore, a factor contributing to the generalization capability of 4DVarNet is its use of a variational cost, effectively ensuring a prior function that captures relevant characteristics of the underlying dynamics, thereby extracting deeper insights from the training data. In other words, it enables the end-to-end architecture to not only perform well in scenarios similar to those it has learned but also to understand the underlying problem, thereby enhancing its ability to adapt to new and diverse situations.

In the considered case-studies for the North Sea, it is important to note that the chosen datasets, though highly effective for offshore areas, may not be ideally suited for shallow water regions like the Dutch Wadden Sea. \FJ{Indeed, the exploited CMEMS L3 products used here are based on satellite remote sensing algorithms (\BBP, \CHL and \SPM), operating notably in the blue-green spectrum, that can be particularly biased in optically shallow waters (waters typically less than 20~m depth, in which light reflected off the seafloor can contribute significantly, when turbidity is low, to the water surface remote sensing reflectance).} However, within the scope of this study, our primary focus is on the interpolation problem – specifically, the task of gap-filling to create gap-free Level 4 (L4) products. While we acknowledge the limitations inherent in our dataset, particularly for shallow waters, we emphasize that our main contribution lies in providing a robust algorithm for interpolation, rather than delving into the complexities of preprocessing in diverse water conditions. And, with that said, our algorithm presents a valuable asset for other teams who focus on data preprocessing. By implementing our method, those teams can first obtain a gap-free L4 product. Following this, they can apply their specialized preprocessing techniques to refine the data further.

\section{Conclusion}\label{sec:conclusion}

In conclusion, this study showcases neural mapping schemes (4DVarNet and UNet) for the space-time interpolation of satellite-derived L3 ocean bio-optical parameters. We demonstrate how neural mapping schemes generalize well across variables and geographical domains, especially 4DVarNet thanks to its design based on a variational data assimilation framework. This can result in a significant reduction of the training costs when considering large-scale case-studies. We also report very significant improvements compared with DInEOF, which is among the-state-of-the-art data-driven interpolation schemes.

For future studies, it would be beneficial to examine the specific scenarios where the neural scheme models excel and where it faces limitations. This exploration could include identifying criteria for selecting the most representative datasets for training the baseline models, which is crucial for effective generalization. 

%It will be important to explore how to enhance the adaptability of the 4DVarNet model to changing conditions, such as evolving environmental factors and data characteristics. This investigation aims to understand and improve the model's resilience and responsiveness to dynamic changes in the datasets it processes, ensuring its continued effectiveness and relevance in varied and shifting contexts.

In addressing real-world implementation challenges, future work should focus on the model's adaptability to a diverse range of variables. Constantly altering the model for each variable is impractical; therefore, developing a model that can generalize across various scenarios is crucial to reduce both training and deployment costs. Additionally, incorporating or studying concepts from model-agnostic or meta-learner in meta-learning, as discussed in \cite{ModelAgnostic2017,LSTM_Metalearning2017}, could significantly enhance the model's generalization capacity, making it more versatile and effective in practical applications.

%\bibliographystyle{IEEEbib}
%\bibliography{strings,refs}

% Nga change the below line to the one same as Clement Fish predict paper
\bibliographystyle{IEEEtran}
\bibliography{biblio}

\end{document}